\documentclass[twocolumn,superscriptaddress]{revtex4-1}
\usepackage{lettrine}
\usepackage{amsmath}
\usepackage{graphicx}
\usepackage{dcolumn}
\usepackage{bm}
\usepackage[colorlinks, linkcolor=blue, citecolor=blue, urlcolor=blue,breaklinks=true]{hyperref}
\usepackage{color}

\begin{document}
\title{Mechanical On-Chip Microwave Circulator}

\author{S. Barzanjeh}
\email{shabir.barzanjeh@ist.ac.at}
\affiliation{Institute of Science and Technology Austria, 3400 Klosterneuburg, Austria}
\author{M. Wulf}
\affiliation{Institute of Science and Technology Austria, 3400 Klosterneuburg, Austria}
\author{M. Peruzzo}
\affiliation{Institute of Science and Technology Austria, 3400 Klosterneuburg, Austria}
\author{M.~Kalaee}
\affiliation{Kavli Nanoscience Institute and Thomas J. Watson, Sr., Laboratory of Applied Physics, California Institute of Technology, Pasadena, CA 91125, USA}
\affiliation{Institute for Quantum Information and Matter, California Institute of Technology, Pasadena, CA 91125, USA}
\author{P.~B.~Dieterle}
\affiliation{Kavli Nanoscience Institute and Thomas J. Watson, Sr., Laboratory of Applied Physics, California Institute of Technology, Pasadena, CA 91125, USA}
\affiliation{Institute for Quantum Information and Matter, California Institute of Technology, Pasadena, CA 91125, USA}
\author{O.~Painter}
\affiliation{Kavli Nanoscience Institute and Thomas J. Watson, Sr., Laboratory of Applied Physics, California Institute of Technology, Pasadena, CA 91125, USA}
\affiliation{Institute for Quantum Information and Matter, California Institute of Technology, Pasadena, CA 91125, USA}
\author{J.~M.~Fink}
\email{jfink@ist.ac.at}
\affiliation{Institute of Science and Technology Austria, 3400 Klosterneuburg, Austria}
\date{\today}
\maketitle
\textbf{Nonreciprocal circuit elements form an integral part of modern measurement and communication systems. Mathematically they require breaking of time-reversal symmetry, typically achieved using magnetic materials~\cite{Jalas2013} and more recently using the quantum Hall effect \cite{Mahoney2017}, parametric permittivity modulation~\cite{Estep2014} or Josephson nonlinearities~\cite{Sliwa2015,Lecocq2017}. Here, we demonstrate an on-chip magnetic-free circulator based on reservoir engineered optomechanical interactions~\cite{Metelmann2015,Toth2017}. Directional circulation is achieved with controlled phase-sensitive interference of six distinct electro-mechanical signal conversion paths. The presented circulator is compact, its silicon-on-insulator platform is compatible with both superconducting qubits~\cite{Keller2017} and silicon photonics, and its noise performance is close to the quantum limit. With a high dynamic range, a tunable bandwidth of up to 30 MHz and an in-situ reconfigurability as beam splitter or wavelength converter \cite{Andrews2015a,Lecocq2016}, it could pave the way for superconducting qubit processors with multiplexed on-chip signal processing and readout.}

Nonreciprocal devices are quintessential tools to suppress spurious modes, interferences and unwanted signal paths. More generally, circulators can be used to realize chiral networks~\cite{Lodahl2017} in systems where directional matter-light coupling is not easily accessible. In circuit quantum electrodynamics %~\cite{Wallraff2004} 
circulators are used for single port coupling or as isolators to protect the vulnerable cavity and qubit states from electromagnetic noise. State of the art passive microwave circulators are based on magneto-optic effects which require sizable magnetic fields incompatible with ultra-low loss superconducting circuits forming a major roadblock towards a fully integrated quantum processor based on superconducting qubits.  

Many recent theoretical and experimental efforts have been devoted to overcome these limitations both in the optical~\cite{Manipatruni2009,Bi2011,KangM.2011%,Lira2012,Tzuang2014,Scheucher2016
} and microwave regimes~\cite{Kamal2011,Abdo2013,Estep2014,Viola2014,%,Abdo2014,Ranzani2014
Sliwa2015,Kerckhoff2015,Lecocq2017,
Mahoney2017}. In parallel, the rapidly growing field of optomechanical and electromechanical systems has shown promising potential for applications in quantum information processing and communication, in particular for microwave to optical conversion~\cite{Barzanjeh2011,Andrews2014} and amplification~\cite{Ockeloen-Korppi2016}. Very recently, several theoretical proposals~\cite{Hafezi2012a,Metelmann2015,Xu2015}
%,Xu2016 
have pointed out that optomechanical systems can lead to nonreciprocity and first isolators have just been demonstrated in the optical domain~\cite{Shen2016,Ruesink2016,Fang2017}. Here, we present an on-chip microwave circulator using a new and tunable silicon electromechanical system.

%%%%%%%%%%%%%%%%%%%%%%%%%
\begin{figure}[t]
\begin{center}
\includegraphics[width=\columnwidth]{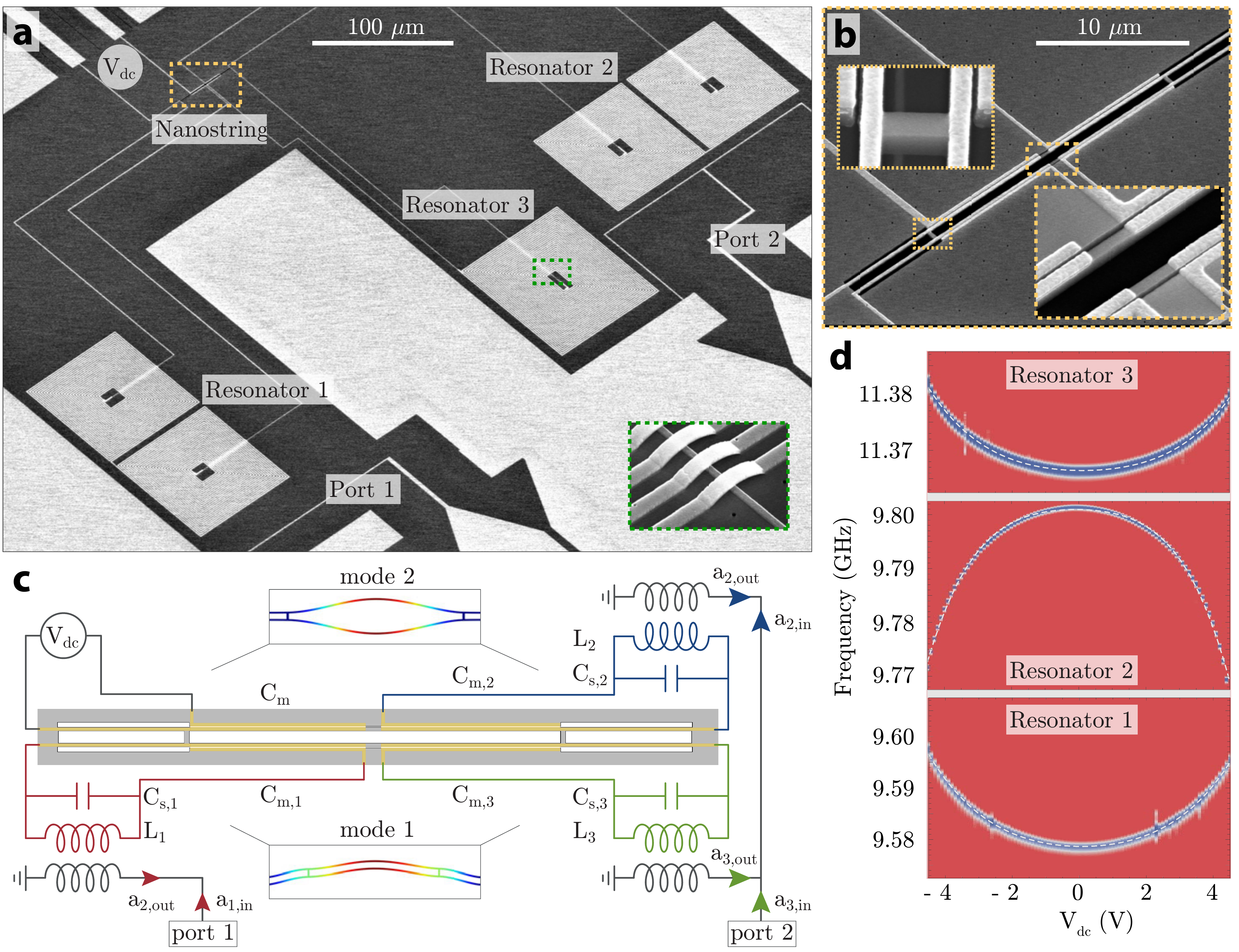}
\caption{
\textbf{Microchip circulator and tunability.} 
\textbf{a}, Scanning electron micrograph of the electromechanical device including three microwave resonators, two physical ports, one voltage bias input ($V_{dc}$) and an inset of the spiral inductor cross-overs (green dashed boxed area). \textbf{b}, Enlarged view of the silicon nanostring mechanical oscillator with four vacuum-gap capacitors coupled to the three inductors and one voltage bias. Insets show details of the nanobeam as indicated by the dashed and dotted rectangles. \textbf{c}, Electrode design and electrical circuit diagram of the device. The input modes $a_{i,in}$ couple inductively to the microwave resonators with inductances $L_i$, coil capacitances $C_i$, additional stray capacitances $C_{s,i}$, and the motional capacitances $C_{m,i}$. The reflected tones $a_{i,out}$ pass through a separate chain of amplifiers each, and are measured at room temperature using a phase locked spectrum analyzer (not shown). The simulated displacement of the lowest frequency in-plane flexural modes of the nanostring are shown in the two insets.~\textbf{d}, Resonator reflection measurement of the three microwave resonators of an identical device, as a function of the applied bias voltage and a fit (dashed lines) to $\Delta\omega\ = \alpha_1 V^2+\alpha_2 V^4$ with the tunabilties $\alpha_1/2\pi=0.53$ MHz/V$^2$ and $\alpha_2/2\pi=0.05$ MHz/V$^4$ with a total tunable bandwidth of 30 MHz for resonator 2 at 9.8 GHz.} 
\label{setup}
\end{center} 
\end{figure}
%%%%%%%%%%%%%%%%%%%%%%%%%%

The main elements of the microchip circulator device are shown in Fig.~\ref{setup} a-b. The circuit is comprised of three high-impedance spiral inductors~($L_i$) capacitively coupled to the in-plane vibrational modes of a dielectric nanostring mechanical resonator. The nanostring oscillator consists of two thin silicon beams that are connected by two symmetric tethers and fabricated from a high resistivity silicon-on-insulator device layer~\cite{Dieterle2016}. Four aluminum electrodes are aligned and evaporated on top of the two nanostrings, forming one half of the vacuum gap capacitors that are coupled to three microwave resonators and one DC voltage bias line as shown schematically in Fig.~\ref{setup}c~(see App.~\ref{circuit} for details). 

The voltage bias line can be used to generate an attractive force which pulls the nanobeam and tunes the operating point frequencies of the device \cite{Andrews2015a}. 
Fig.~\ref{setup}~d shows the measured resonance frequency change as a function of the applied bias voltage $V_{dc}$. As expected, resonators 1 and 3 are tuned to higher frequency due to an increased vacuum gap while resonator 2 is tuned to lower frequency. A large tunable bandwidth of up to $30$~MHz as obtained for resonator 2, the ability to excite the motion directly and to modulate the electromechanical coupling in-situ represents an important step towards new optomechanical experiments and more practicable on-chip reciprocal and nonreciprocal devices. 

As a first step we carefully calibrate and characterize the individual electromechanical couplings and noise properties. %, similar to Ref.~\cite{Fink2016}. 
We then measure the bidirectional frequency conversion between two microwave resonator modes as mediated by one mechanical mode~\cite{Lecocq2016}. 
The incoming signal photons can also be distributed to two ports with varying probability as a function of the parametric drive strength and in direct analogy to a tunable beam splitter. We present the experimental results, the relevant sample parameters and the theoretical analysis of this bidirectional frequency conversion process in App. \ref{wavelengthCon}. 

\begin{figure}[t]
\centering
\includegraphics[width=0.65\columnwidth]{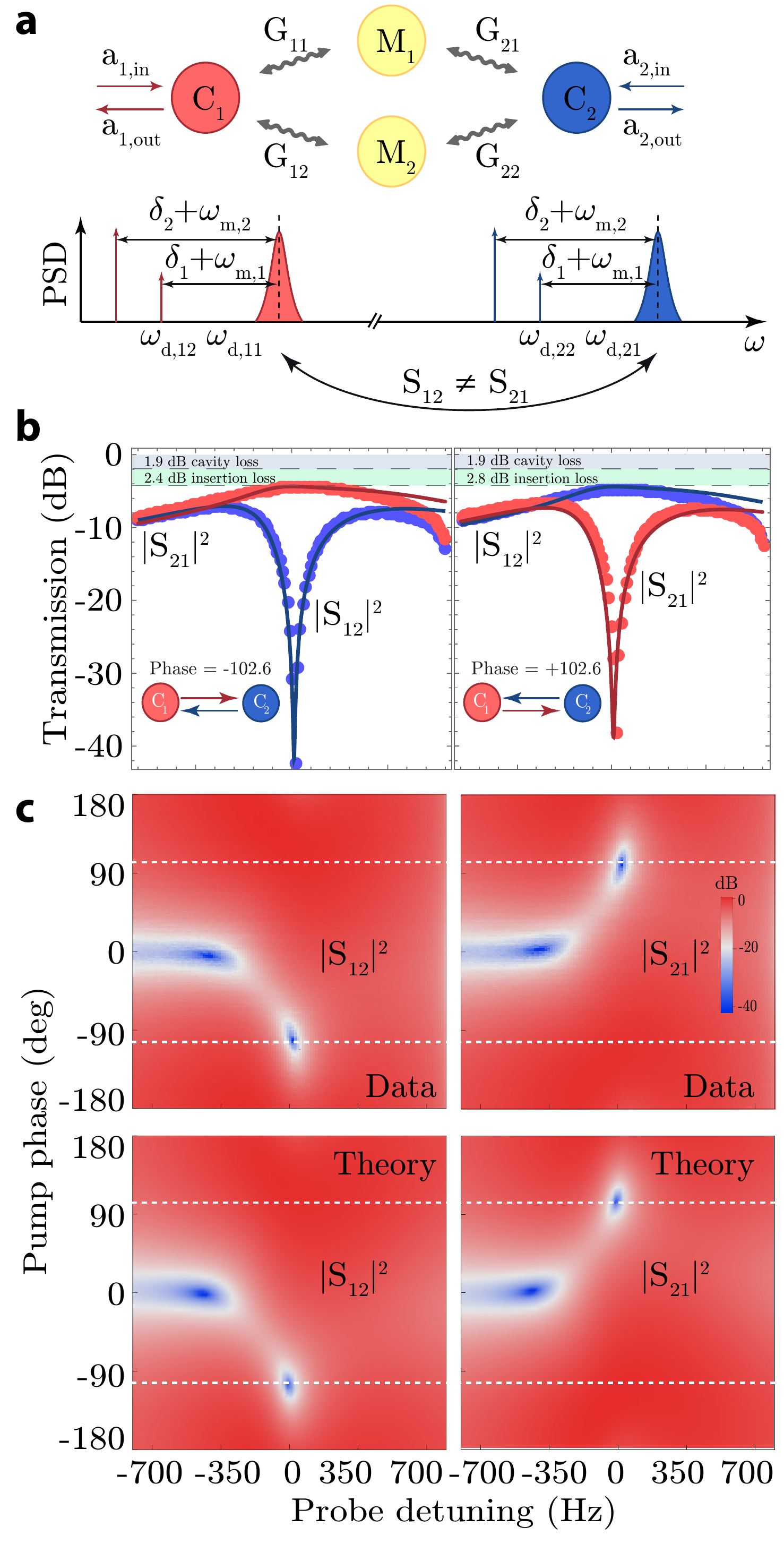}
\caption{\textbf{Optomechanical isolator.} 
\textbf{a}, Mode coupling diagram for optomechanically induced nonreciprocity. Two microwave cavities~($C_1$ and $C_2$) are coupled to two mechanical modes ($M_1$ and $M_2$) with the optomechanical coupling rates $G_{ij}$ (where $i,j=1,2$), inducing two distinct signal conversion paths. Power spectral density (PSD) of the two microwave cavities and arrows indicating the frequency of the four microwave pump tones slightly detuned by $\delta_i$ from the lower motional sidebands of the resonances. All four pumps are phase-locked while the signal tone is applied. Only one of the microwave source phases is varied to find the optimal interference condition for directional transmission between port 1 and 2. 
\textbf{b}, Measured power transmission~(dots) in forward $|S_{21}|^2$ (cavity 1 $\rightarrow$ cavity 2) and backward directions $|S_{12}|^2$ (cavity 2 $\rightarrow$ cavity 1) as a function of probe detuning for two different phases $\phi=\pm 102.6$ degrees. The solid lines show the results of the coupled-mode theory model discussed in the text. 
\textbf{c}, Experimental data (top) and theoretical model (bottom) of measured transmission coefficients $|S_{12}|^2$ and  $|S_{21}|^2$ as a function of signal detuning and pump phase $\phi$. Dashed-lines indicate the line plot locations of panel b.} \label{Isolator}
\end{figure}

Directionality is achieved by engaging the second mechanical mode, a method which was developed in parallel to this work~\cite{Bernier2016,Peterson2017} for demonstrating nonreciprocity in single-port electromechanical systems. We begin with the theoretical model describing two microwave cavities with resonance frequencies $\omega_i$ and total linewidths $\kappa_i$ with $i=1,2$ parametrically coupled to two distinct modes of a mechanical resonator with resonance frequencies $\omega_{m,j}$ and damping rates $\gamma_{m,j}$ with $j=1,2$. To establish the parametric coupling, we apply four microwave tones, with frequencies detuned by $\delta_j$ from the lower motional sidebands of the resonances, as shown in Fig. \ref{Isolator}a. In a reference frame rotating at the frequencies $\omega_i$ and $\omega_{m,j}+\delta_j$, the linearized Hamiltonian in the resolved sideband regime ($\omega_{m,j} \gg \kappa_1,\kappa_2$) is given by ($\hbar=1$)
\begin{eqnarray}\label{isolatorHam}
H=-\sum_{j=1,2} \delta_j b_j^{\dagger}b_j&+&\sum_{i,j=1,2}G_{ij}\Big(e^{i\phi_{ij}}a_ib^{\dagger}_j+
e^{-i\phi_{ij}}a^{\dagger}_ib_j\Big)\nonumber\\
&&+H_{\mathrm{off}},
\end{eqnarray}
where $a_{i}\,(b_j)$ is the annihilation operator for the cavity $i$~(mechanics $j$), $G_{ij}=g_{0,ij}\sqrt{n_{ij}}$ and  $g_{0ij}$ are the effective and vacuum electromechanical coupling rates between the mechanical mode $j$ and cavity $i$ respectively, while $n_{ij}$ is the total number of photons inside the cavity $i$ due to the drive with detuning $\Delta_{ij}$, and $\phi_{ij}$ is the relative phase set by drives. Here, $\Delta_{11}=\Delta_{21}=\omega_{m,1}+\delta_1$ and $\Delta_{22}=\Delta_{12}=\omega_{m,2}+\delta_2$ are the detunings of the drive tones with respect to the cavities and $H_{\mathrm{off}}$ describes the time dependent coupling of the mechanical modes to the cavity fields due to the off-resonant drive tones. These additional coupling terms create cross-damping \cite{Buchmann2015} and renormalize the mechanical modes, and can only be neglected in the weak coupling regime for $G_{ij},\kappa_j\ll\omega_j, |{\omega_{m,2}-\omega_{m,1}}|$.

%%%%%%%%%%%%%%%%%%%%%%%%%%%
%%%%%%%%%%%%%%%%%%%%%%%%%%%

To see how the nonreciprocity arises we use the quantum Langevin equations of motion along with the input-output theorem to express the scattering matrix $S_{ij}$ of the system described by the Hamiltonian (\ref{isolatorHam}), and relating the input photons $a_{\mathrm{in},i}(\omega_i)$ at port $i$ to the output photons $a_{\mathrm{out},j}(\omega_j)$ at port $j$ via $a_{\mathrm{out},i}=\sum_{j=1,2}S_{ij}a_{\mathrm{in},i}$ with $i=1,2$. The dynamics of the four-mode system described by Hamiltonian (\ref{isolatorHam}) is fully captured by a set of linear equations of motion as verified in App.~\ref{TheoryModel}. Solving these equations in the frequency domain, using the input-output relations, and setting $\phi_{22}=\phi,$ $\phi_{11}=\phi_{21}=\phi_{12}=0$, the ratio of backward to forward transmission reads
\begin{equation}\label{nonreciprocity}
\lambda:=\frac{S_{12}(\omega)}{S_{21}(\omega)}=\frac{\sqrt{C_{11}C_{21}}\Sigma_{m,2}(\omega)+\sqrt{C_{12}C_{22}}\Sigma_{m,1}(\omega)e^{i\phi}}{\sqrt{C_{11}C_{21}}\Sigma_{m,2}(\omega)+\sqrt{C_{12}C_{22}}\Sigma_{m,1}(\omega)e^{-i\phi}}.
\end{equation}
Here, $\Sigma_{m,j}=1+2i\big[(-1)^j\delta-\omega\big]/\gamma_{m,j}$ is the inverse of the mechanical susceptibility divided by the mechanical linewidth $\gamma_{m,j}$ and $C_{ij}=4G_{ij}^2/(\kappa_i \gamma_{m,j})$ is the optomechanical cooperativity. Note that, in Eq.~(\ref{nonreciprocity}) we assume the device satisfies the impedance matching condition on resonance i.e. $S_{ii}(\omega=0)=0$ which can be achieved in the high cooperativity limit ($C_{ij}\gg 1$).

Inspection of equation~(\ref{nonreciprocity}) reveals the crucial role of the relative phase between the drive tones $\phi$ and the detuning $\delta$ to obtain nonreciprocal transmission. When the cooperativities for all four optomechanical couplings are equal~($C_{ij}=\mathcal{C}$) then perfect isolation, i.e.~$\lambda = 0$, occurs for
\begin{equation}\label{phase}
\mathrm{tan}[\phi(\omega)]=\frac{\delta(\gamma_{m,1}+\gamma_{m,2})+\omega(\gamma_{m,2}-\gamma_{m,1})}{\gamma_{m,1}\gamma_{m,2}/2-2(\delta^2-\omega^2)}.
\end{equation}
Equation \ref{phase} shows that on resonance ($\omega=0$) $\mathrm{tan}[\phi] \propto \delta$, highlighting the importance of the detuning $\delta$ to obtain nonreciprocity. Tuning all four drives to the exact red sideband frequencies ($\delta=0$) results in bidirectional behavior ~($\lambda=1$). At the optimum phase $\phi$ given by Eq.~(\ref{phase}), $\omega=0$, and for two mechanical modes with identical decay rates~($\gamma_{m,1}=\gamma_{m,2}=\gamma$) the transmission in forward direction is given by
\begin{eqnarray}\label{trans}
S_{21}=-\sqrt{\eta_1\eta_2}\left[
\frac{4i\,\delta(1-2i\delta/\gamma)}{\mathcal{C}\gamma\left(1+\frac{1+4\delta^2/\gamma^2}{2\mathcal{C}}\right)^2}\right]
\end{eqnarray}
where $\eta_{1(2)}=\kappa_{\mathrm{ext},{1(2)}}/\kappa_{1(2)}$ is the resonator coupling ratio and $\kappa_i=\kappa_{\mathrm{int},i}+\kappa_{\mathrm{ext},i}$ is the total damping rate.  Here $\kappa_{\mathrm{int},i}$ denotes the internal loss rate and $\kappa_{\mathrm{ext},i}$ the loss rate due to the cavity to waveguide coupling. Equation~(\ref{trans}) shows that the maximum of the transmission in forward direction, $|S_{21}|^2=\eta_1\eta_2[1-(2\mathcal{C})^{-1}]$, occurs when $2\mathcal{C}=1+4\delta^2/\gamma^2$ and for large cooperativities $\mathcal{C}\gg 1$. These conditions, as implemented in our experiment, enable the observation of asymmetric frequency conversion with strong isolation in the backward direction and small insertion loss in forward direction. 

Using the on-chip electromechanical microwave circuit shown in Fig.~\ref{setup}~a, we experimentally realize directional wavelength conversion between two superconducting coil resonators at $(\omega_1,\omega_2)/2\pi=(9.55,9.82)$~GHz coupled to two different physical waveguide ports and measurement lines with $(\eta_1,\eta_2)=(0.74,0.86)$. Here, we use the two lowest-frequency vibrational in-plane modes of the mechanical resonator at $(\omega_{m,1},\omega_{m,2})/2\pi=(4.34,5.64)$~MHz with intrinsic damping rates $(\gamma_{m,1},\gamma_{m,2})/2\pi=(4,8)$~Hz. The vacuum optomechanical coupling strengths for these mode combinations are $(g_{0,11},g_{0,12},g_{0,21},g_{0,22})/2\pi=(33,34,13,31)$~Hz. 

Figure \ref{Isolator}~b shows the measured transmission of the wavelength conversion in the forward $|S_{21}|^2$ and backward directions $|S_{12}|^2$ as a function of probe detuning for two different phases as set by one out of the four phase locked microwave drives. At $\phi=-102.6$ degree and over a frequency range of $1.5$~kHz we measure high transmission from cavity 1 to 2 with an insertion loss of $2.4$~dB while in the backward direction the transmission is suppressed by up to $40.4$~dB. Likewise, at the positive phase of $\phi=102.6$ degree the transmission from cavity 1 to 2 is suppressed while the transmission from cavity 2 to 1 is high. In both cases we observe excellent agreement with theory (solid lines). Fig.~\ref{Isolator}~c shows the S parameters for the whole range of phases $\phi$, which are symmetric and bidirectional around $\phi=0$. We find excellent agreement with theory over the full range of measured phases with less than $10\%$ deviation to independently calibrated drive photon numbers and without any other free parameters.

For bidirectional wavelength conversion, higher cooperativity enhances the bandwidth. In contrast, the bandwidth of the nonreciprocal conversion is independent of cooperativity and set only by the intrinsic mechanical linewidths $\gamma_{m,i}$, which can be seen in Eq~(\ref{nonreciprocity}). This highlights the fact that the isolation appears when the entire signal energy is dissipated in the mechanical environment, a lossy bath that can be engineered effectively~\cite{Toth2017}. In the present case it is the off-resonant coupling between the resonators and the mechanical oscillator which modifies this bath. The applied drives create an effective interaction between the mechanical modes, where one mode acts as a reservoir for the other and vice versa. This changes both the damping rates and the eigenfrequencies of the mechanical modes. %\cite{Buchmann2015}.
%,Noguchi2016
It therefore increases the instantaneous bandwidth of the conversion and automatically introduces the needed detuning,
%of $\delta_i=XXX$
%. The cross-coupling depends only on the optomechanical coupling rates and the mechanical frequencies 
which is fully taken into account in the theory.
%and fits.% without free parameters.

\begin{figure}[t]
\centering
\includegraphics[width=\columnwidth]{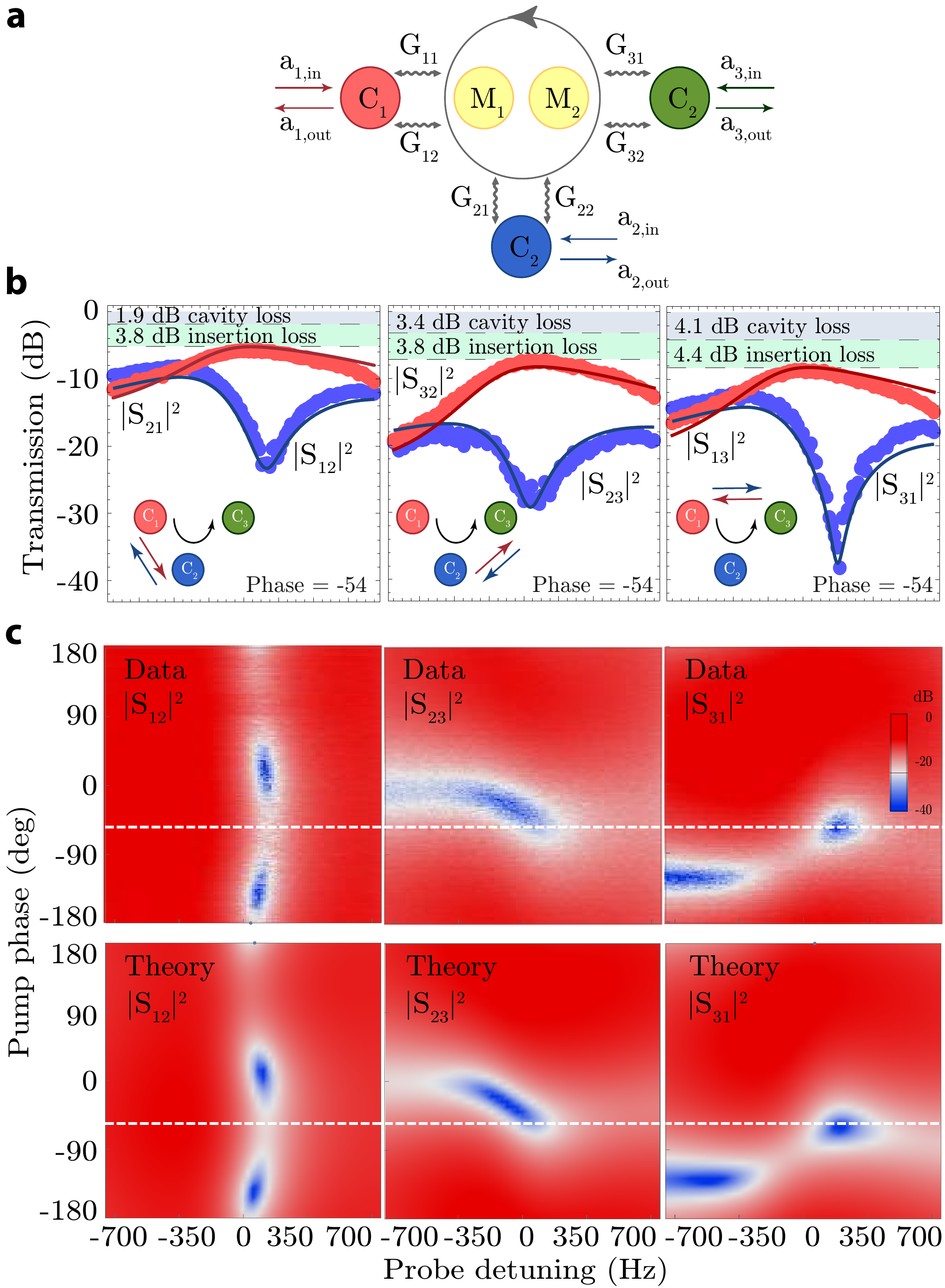}
\caption{\textbf{Optomechanical circulator.} \textbf{a}, Mode coupling diagram describing the coupling between three microwave cavities~($C_1$, $C_2$ and $C_3$) and two mechanical modes ($M_1$ and $M_2$) with optomechanical coupling rates $G_{ij}$ (where $i=1,2,3 $ and $j=1,2$), creating a circulatory frequency conversion between the three cavity modes. \textbf{b},  Measured power transmission~(dots) in forward ($|S_{21}|^2$, $|S_{32}|^2$ and $|S_{13}|^2$) and backward directions ($|S_{12}|^2$, $|S_{23}|^2$ and $|S_{31}|^2$) as a function of probe detuning for a pump phase $\phi=-54$ degrees. The solid lines show the prediction of the coupled-mode theory model discussed in the text. \textbf{c}, Measured S parameters (top) and theoretical model (bottom) as a function of detuning and pump phase. Dashed-lines indicate the line plot positions shown in panel \textbf{b}.} \label{circulator}
\end{figure}

The described two-port isolator can be extended to an effective three-port device by parametrically coupling the third microwave resonator capacitively to the dielectric nanostring, as shown in Fig.~\ref{setup} a. The third resonator at a resonance frequency of $\omega_3/2\pi=11.30$~GHz is coupled to the waveguide with $\eta_3=0.52$ and to the two in-plane mechanical modes with $(g_{0,31},g_{0,32})/2\pi=(22,45)$~Hz. Similar to the isolator, we establish a parametric coupling between cavity and mechanical modes using six microwave pumps with frequencies slightly detuned from the lower motional sidebands of the resonances, which for certain pump phase combinations can operate as a three-port circulator for microwave photons, see Fig~\ref{circulator}~a.  Using an extra microwave source as probe signal, we measure the power transmission between all ports and directions as shown in Fig.~\ref{circulator} b for a single fixed phase of $\phi=-54$ degree, optimized experimentally for forward circulation. 

At this phase we see high transmission in the forward direction $S_{21,32,13}$ with an insertion loss of ($3.8$, $3.8$, $4.4$)~dB and an isolation in the backward direction $S_{12,23,31}$ of up to ($18.5$, $23$, $23$)~dB. The full dependence of the circulator scattering parameters on the drive phase is shown in Fig.~\ref{circulator}~c where we see excellent agreement with theory. The added noise photon number of the device is found to be $(n_{\mathrm{add},21},n_{\mathrm{add},32},n_{\mathrm{add},13})=(4,6.5,3.6)$ in the forward direction and $(n_{\mathrm{add},12},n_{\mathrm{add},23},n_{\mathrm{add},31})=(4,4,5.5)$ in the backward direction, limited by the thermal occupation of the mechanical modes and discussed in more detail in App.~\ref{addnoise}.

In conclusion, we realized a frequency tunable microwave isolator / circulator that is highly directional and operates with low loss and added noise. Improvements of the circuit properties will help increase the instantaneous bandwidth and further decrease the transmission losses of the device. The external voltage bias offers new ways to achieve directional amplification and squeezing of microwave fields in the near future. Direct integration with superconducting qubits should allow for on-chip single photon routing as a starting point for more compact circuit QED experiments.  

\textbf{Acknowledgements} We thank Nikolaj Kuntner for the development of the Python virtual instrument panel and Georg Arnold for supplementary device simulations. This work was supported by IST Austria and the European Union's Horizon 2020 research and innovation program under grant agreement No 732894 (FET Proactive HOT). SB acknowledges support from the European Union's Horizon 2020 research and innovation program under the Marie Sklodowska Curie grant agreement No 707438 (MSC-IF SUPEREOM).
\\
\noindent\textbf{Author contributions}
SB and JMF conceived the ideas for the experiment. SB developed the theoretical model, performed and analyzed the measurements. SB, MW, MP and JMF designed the microwave circuit and built the experimental setup. MK, PD, JMF and OP designed the mechanical nanobeam oscillator. JMF and MK fabricated the sample. PD and OP contributed to sample fabrication. SB and JMF wrote the manuscript. JMF supervised the research.
\\
\noindent\textbf{Additional information}
Correspondence and requests for materials should be addressed to SB and JMF.
\\
\textbf{Competing financial interests}
The authors declare no competing financial interests.

%%%%%%%%%%%%%%%%%%%%%%%%%%%%%%%%
%%%%%%%%%%%%%%%%%%%%%%%%%%%%%%%%

\clearpage
\begin{widetext}
\section{Supplementary Information}
\appendix
\section{Circuit properties}\label{circuit}
The electromechanical microwave circuit shown in Fig.~1~a, includes three high-impedance microwave spiral inductors~($L_i$) capacitively coupled to the in-plane vibrational modes of a dielectric nanostring mechanical resonator, creating three LC resonators with frequencies~$\omega_i=1/\sqrt{L_i C_i}$ with $i=1,2,3$. The nanostring resonator fabricated from a high resistivity smart-cut silicon-on-insulator wafer with $220$~nm device layer thickness has a length of $9.4\,\mu$m and consists of two metalized beams that are connected with two tethers at their ends. The vacuum gap size for the mechanically compliant capacitor fabricated with an inverse shadow technique \cite{Pitanti2015} is approximately $60$~nm. 

The electromechanical coupling between the nanostring mechanical resonator and each LC circuit is given by 
\begin{equation}\label{g0equation}
g_{0i}=x_{\mathrm{zpf}}\frac{\partial \omega_i}{\partial v}=-x_{\mathrm{zpf}}\zeta_i\frac{\omega_i}{2C_{m,i}}\frac{\partial C_{m,i}}{\partial v},
\end{equation}
where $v$ is the amplitude coordinate of the in-plane mode, $\zeta_i=\frac{C_{m,i}}{C_{\Sigma,i}}$ is the participation ratio
of the vacuum gap capacitance~$C_{m,i}$ to the total capacitance of the circuit $C_{\Sigma,i}=C_{m,i}+C_{s,i}$, where $C_{s,i}$ is the stray capacitance of the circuit including the intrinsic self-capacitance of the inductor
coils. Eq.~(\ref{g0equation}) indicates that large electromechanical coupling $g_{0i}$ requires a large participation ratio. We can make the coil capacitance $C_{L,i}$ relatively small by using a suspended and tightly wound rectangular spiral inductor
with a wire width of $500$~nm and wire-to-wire
pitch of $1\mu$m \cite{Fink2016}. Knowing the inductances $L_i$ of the fabricated inductors based on modified Wheeler, as well as the actually measured resonance frequencies $\omega_i$ along with  vacuum-gap capacitance $C_m$ (from FEM simulations), we can find the total stray capacitance including the intrinsic self-capacitance of the each inductor coil correspondingly. Careful thermometry calibrated mechanical noise spectroscopy measurements similar to the ones in \cite {Fink2016} yield the measured electromechanical coupling for each mode combination as outlined in the table below.
\begin{center}\label{cavitytable}
    \begin{tabular}{| l | l | l | l | l | l | l | l | l | l | l |}
    \hline
     & $\frac{\omega}{2\pi}$(GHz) & $\frac{\kappa_{\mathrm{int}}}{2\pi}$(MHz) & $\frac{\kappa_{\mathrm{ex}}}{2\pi}$(MHz) & $\frac{\kappa}{2\pi}$ (MHz) & $\eta=\frac{\kappa_{\mathrm{ex}}}{\kappa}$ & $L$(nH) & $C_{s}$(fF) & $C_m$(fF) &  $\frac{g_{01}}{2\pi}$(Hz) &  $\frac{g_{02}}{2\pi}$(Hz)  \\ \hline
    Cavity 1 & 9.55 & 0.62 & 1.8 & 2.42 & 0.74 & 48.2 & 5.3 & 0.45 & 33 &34 \\ \hline
    Cavity 2 & 9.82 & 0.28 & 1.7 & 1.98 & 0.86 & 48.3 & 4.98  & 0.45& 13& 31\\ \hline
    Cavity 3 & 11.32 & 1.42 & 1.58 & 3 & 0.52 & 34.4 & 5.29 & 0.45& 22& 45  \\
    \hline
\end {tabular}
\end{center}

We use finite-element method (FEM) numerical simulations to find the relevant in-plane mechanical modes of the structure and optimize their zero point displacement amplitudes and mechanical quality factor. Our simulations are consistent with the measured mechanical frequencies for a tensile stress of $\sim$600~MP in a $\sim$70~nm thick electron beam evaporated aluminum layer \cite{Regal2008}. The associated effective mass and zero-point displacement amplitude along with the measured linewidths and resonance frequencies of the first two in-plane modes of the nanostring are presented in the table below.
\begin{center}\label{mechanicstable}
    \begin{tabular}{| l | l | l | l | l |}
    \hline
     & $\frac{\omega_m}{2\pi}$(MHz) & $\frac{\gamma_m}{2\pi}$(Hz) & $m_{\mathrm{eff}}$(pg) & $x_{\mathrm{zpf}}$ (fm) \\ \hline
   first mechanical mode & 4.34 & 4 & 4 & 22 \\ \hline
   second mechanical mode & 5.64 & 8 & 2.2 & 26 \\ \hline
  \end{tabular}
\end{center}

%%%%%%%%%%%%%%%%%%%%%%%%%%%%%%%%%

\section{Bidirectional frequency conversion}\label{wavelengthCon}
To understand the optomechanical frequency conversion, we first theoretically model our system to see how frequency conversion arises. Figure~\ref{FigWLC}~a shows an electromechanical system, in which two microwave cavities with resonance frequencies $\omega_{1}$ and $\omega_2$ and linewidths $\kappa_{1}$ and $\kappa_2$ are coupled
to a mechanical oscillator with frequency $\omega_m$ and damping rate~$\gamma$. The electromechanical coupling is driven by two strong drive fields, $\mathcal{E}_1$ and $\mathcal{E}_2$, near the red sideband of the respective microwave modes at $\omega_{d,1(2)}=\omega_{1(2)}-\omega_m$, see Fig.~\ref{FigWLC}~b. In the resolved-sideband limit~($\omega_m\gg \kappa_{1(2)},\gamma$) the linearized electromechanical Hamiltonian in the rotating frames with respect to the external driving fields is given by~($\hbar=1$)
\begin{equation}\label{hamiltonian1}
H=\sum_{i=1,2}\Delta_i a^{\dagger}_i a_i+\omega_m b^{\dagger} b+\sum_{i=1,2}G_i\Big(a_ib^{\dagger}+b a_i^{\dagger}\Big),
\end{equation}
where $a_{1(2)}$ is the annihilation operator for the microwave signal field 1 (microwave signal field 2), $b$ is the annihilation operator of the mechanical mode, $\Delta_{1(2)}=\omega_{1(2)}-\omega_{d1(2)}=\omega_m$ is the detuning between the external driving field and the relevant cavity
resonance, and $G_{i}=g_{0i}\sqrt{n_i}$ is the effective electromechanical coupling rate between the mechanical resonator and cavity $i$ with $n_{i}=\frac{2\mathcal{E}_i}{\kappa_{i}^2+4\Delta_i^2} $ being the total number of photons inside the cavity. Note that, the fast-oscillating counter-rotating terms at $\pm2\omega_m$ are omitted from the Hamiltonian under the rotating wave approximation. 

\begin{figure}[t]
\centering
\includegraphics[width=0.35\columnwidth]{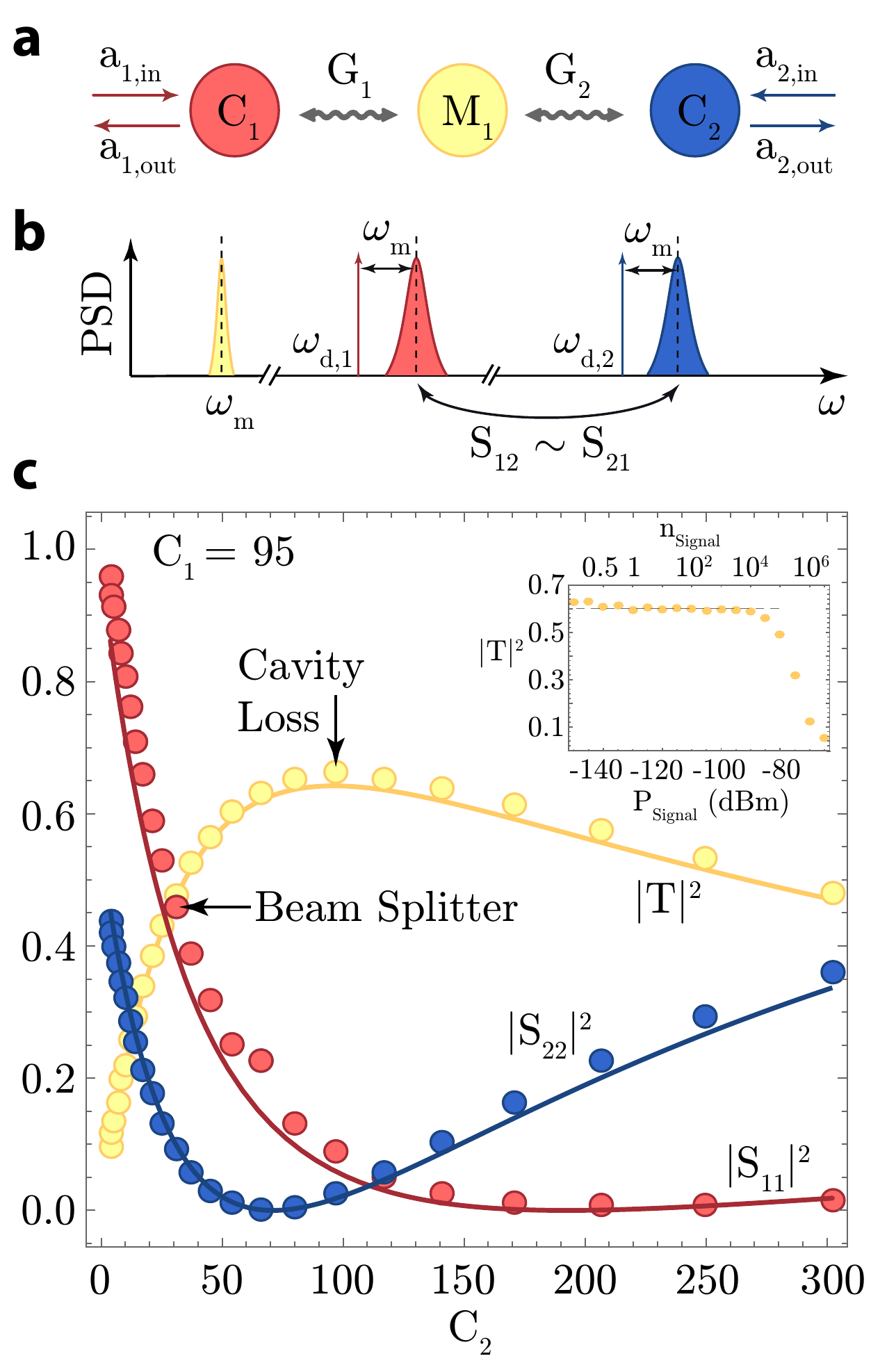}
\caption{\textbf{Bidirectional frequency conversion}. \textbf{a}, The microwave-mechanical mode diagram for the frequency conversion. Two microwave cavities $C_1$ and $C_2$ are parametrically coupled to a mechanical mode with coupling rates $G_{1}$ and $G_{2}$, which gives rise to frequency conversion between the two microwave cavities. \textbf{b}, Power spectral densities~(PSD) of the mechanical mode  and microwave cavities and the drive tone frequencies indicated with vertical arrows near the red sidebands of the microwave modes at $\omega_{d,1(2)}=\omega_{1(2)}-\omega_m$. \textbf{c}, Experimental demonstration~(dots) and theoretical prediction~(solid lines) of the frequency conversion between two microwave cavities at resonance frequencies ~$(\omega_1,\omega_2)/2\pi=(9.55,9.82)$~GHz as a function of cooperativity $C_2$ for $C_1=95$. Here, $|T|^2=|S_{12}|\cdot |S_{21}|$ (yellow dots), $|S_{11}|^2 $ (red dots) and  $|S_{22}|^2$ (blue dots) demonstrate the magnitude of the transmission and reflection coefficients on resonance with the cavities, respectively. As predicted by Eq.~(\ref{transmission}), the transmission between the two cavities is maximum for $C_1=C_2 \approx 95$. The inset shows the dynamic range of the device where the transmission coefficient is measured as function of the signal input power $P_{\mathrm{signal}}$ or mean total number of signal photons inside the microwave cavities $n_{\mathrm{signal}}$.} \label{FigWLC}\end{figure}

The first and second terms of Hamiltonian~(\ref{hamiltonian1}) describe the free energy of the mechanical and cavity modes while the last term of the Hamiltonian indicates a beam splitter-like interaction between mechanical degree of freedom and microwave cavity modes. In fact this term allows both optomechanical cooling (with rate $\Gamma_i=4G_i^2/\kappa_i$) and  bidirectional photon conversion between two distinct microwave frequencies. In the photon conversion process, first an input microwave signal at frequency~$\omega_1$ with amplitude $a_{\mathrm{in},1}(\omega_1)$ is down-converted into the mechanical mode at frequency $\omega_m$, i.e.~$a_1(\omega_1)\xrightarrow{H \propto a_1b^{\dagger}} b(\omega_m)$. 
%This process corresponds to $a^{\dagger}_1b$ in Hamiltonian~(\ref{hamiltonian1}). 
Next, during an up-conversion process the mechanical mode transfers its energy to the output of the other microwave cavity at frequency~$\omega_2$ and amplitude $a_{\mathrm{out},2}(\omega_2)$, i.e.~$b(\omega_m)\xrightarrow{H \propto b a^{\dagger}_2} a_2(\omega_2)$. 
%This process can be explained by $b^{\dagger}a_2$ in Hamiltonian~(\ref{hamiltonian1}). 
Likewise, an input microwave signal at frequency~$\omega_2$ can be converted to frequency~$\omega_1$ by reversing the conversion process. In fact, the Hermitian aspect of the Hamiltonian~(\ref{hamiltonian1}) makes the conversion process bidirectional and holds the time-reversal symmetry.

The photon conversion efficiency, which is defined as the ratio of the output-signal photon flux over the input-signal photon flux, is given by $|S_{21}|^2=\Big|\frac{a_{\mathrm{out},2(\omega_2)}}{a_{\mathrm{in},1(\omega_1)}}\Big|^2$. Since the conversion process is bidirectional therefore  $|S_{21}|=|S_{12}|=|T|$. In the steady state and in the weak coupling regime the conversion efficiency reduces to 
\begin{equation}\label{transmission}
|T|^2=\frac{4\eta_1\eta_2 C_1C_2}{(1+C_1+C_2)^2},
\end{equation}
where $C_{1(2)}=\frac{4g_{0,1(2)}^2n_{1(2)}}{\kappa_{1(2)}\gamma_m}$ is the electromechanical cooperativity for cavity 1 (2) and $\eta_{1(2)}=\frac{\kappa_{\mathrm{ext},{1(2)}}}{\kappa_{1(2)}}$ is the output coupling ratio in which $\kappa_i=\kappa_{\mathrm{int},i}+\kappa_{\mathrm{ext},i}$ is the total damping rate while $\kappa_{\mathrm{int},i}$ and $\kappa_{\mathrm{ext},i}$ show the intrinsic and extrinsic decay rate of the microwave cavities, respectively. Likewise, the reflection coefficients due to impedance mismatch are given by
\begin{eqnarray}\label{transmission2}
|S_{11}|^2&=&\Big(\frac{1+C_1+C_2-2\eta_1(1+C_2)}{1+C_1+C_2}\Big)^2,\\
|S_{22}|^2&=&\Big(\frac{1+C_1+C_2-2\eta_2(1+C_1)}{1+C_1+C_2}\Big)^2.
\end{eqnarray}

Note that for the lossless microwave cavities~($\eta_i=1$), near unity photon conversion can be achieved in the limit that $C_1=C_2=C$ and $C \gg1$. The former condition balances the photon-phonon conversion rate for each cavity while the later condition guarantees the mechanical damping rate $\gamma_m$ is much weaker than the damping rates $\Gamma_{i}=\gamma_m C_{i}$. 
%Therefore conversion process is capable of exchanging photons coherently such that the photon-to-photon conversion predominates mechanical damping rate $\gamma_m$ (the rate at which a single phonon is exchanged with the oscillator’s environment). 
Under these two conditions, the ideal photon conversion is achieved i.e.~$|T|^2=1$ (perfect transmission) and $|S_{11}|^2=|S_{22}|^2=0$ (no reflection). The denominator of Eq.~(\ref{transmission}) indicates that the bandwidth of the conversion is given by $\Gamma_T=\gamma_m+\Gamma_1+\Gamma_2$, which is the total back-action-damped linewidth of the mechanical resonator in the presence of the two microwave drive fields.

We perform coherent microwave frequency conversion using the intermediate nanostring resonator as a coupling element between two superconducting coil resonators at $\omega_1/2\pi=9.55$~GHz and $\omega_2/2\pi=9.82$~GHz as shown in Fig~\ref{setup}~a.  
%The device anchored to the mixing chamber of a cryogen free dilution refrigerator with a fridge temperature of $T_f=10$mK and all incoming lines are heavily filtered and attenuated to eliminate Johnson and phase noises. 
The microwave cavities are accessible by “ports", i.e.~semi-infinite transmission lines giving the modes finite energy decay rates leading to the cavity linewidths $\kappa_1/2\pi=2.42$ MHz and $\kappa_2/2\pi=1.98$ MHz with associated output coupling ratios $\eta_1=0.74$ and $\eta_2=0.86$, indicating that both cavities are strongly overcoupled to the two distinct physical ports 1 and 2. The fundamental mode of the mechanical oscillator has a resonance frequency of $\omega_m/2\pi=4.34$MHz with the corresponding damping rate of $\gamma_m/2\pi=4$Hz. Measuring the mechanical resonator noise spectrum along with the off-resonant reflection coefficients of each cavity and measurement line, we calibrate the gain and attenuation in each input-output line and accurately back out the vacuum optomechanical coupling rate for each cavity of $g_{01}/2\pi=33$Hz and $g_{02}/2\pi=13$Hz. 

Figure \ref{FigWLC}~c shows the measured scattering parameters $|S_{11}|^2$ (red  line), $|S_{22}|^2$ (blue line), and $|T|^2=|S_{12}|\cdot|S_{21}|$ (yellow line) versus the electromechanical cooperativity
 $C_2$ at $C_1=95$. As predicted by Eq.~(\ref{transmission}) at $C_1=C_2\simeq 95$ we measure a transmission of $|T|^2=0.64$, which is dominated by internal losses of the cavities limiting the maximum reachable conversion efficiency to $|T|^2\leq\eta_1\eta_2=0.64$.
 
Another important aspect of such a transducer is the dynamic range of the device. In the inset of Fig.~\ref{FigWLC}~c we show measured maximum transmission as a function of the applied signal power. Our results demonstrate that high conversion efficiencies can be maintained up to about $-80$~dBm input signal power, corresponding to about $10^5$ signal photons inside the cavities. At even higher signal powers the transmission efficiency is degraded abruptly, because the probe tone acts as an additional strong drive invalidating the transducer model, and also because of an increase of the resonance frequency shifts and resonator losses. 

%%%%%%%%%%%%%%%%%%%%%%%%%%%%%%%%%%%%%%%%%

\section{General theory of a coupled electromechanical system}\label{TheoryModel}
%%%%%%%%%%%%%%

\subsection{Hamiltonian of a multi-mode electromechanical transducer}
In this section we present a general theory to describe the nonreciprocal behavior of our on-chip electromechanical transducer, shown in Fig~1a of the main paper. We begin with an optomechanical system comprised of three microwave cavities with frequencies $\omega_i$ and linewidths $\kappa_i$ where $i=1,2,3$ that are coupled to two vibrational modes of a mechanical oscillator with frequencies $\omega_{m.i}$ and damping rates $\gamma_{m,i}$ where $i=1,2$.
%,see Fig~\ref{circulator}~a. 
To tune a desired coupling into resonance, we assume the cavities are coherently driven with six microwave tones, with frequencies detuned from the lower motional sidebands of the resonances by $\delta_{0,i}$. The Hamiltonian of the system is~($\hbar=1$)\cite{Barzanjeh2015}
\begin{equation}\label{Ham1}
H=\sum_{i=1}^{3}\omega_ia_i^{\dagger}a_i+\sum_{i=1}^2\omega_{m,i}b_i^{\dagger}b_i+\sum_{i=1}^3\sum_{j=1}^2g_{0,ij}a_i^{\dagger}a_i(b_j+b_j^{\dagger})+H_d,
\end{equation}
where $a_{i}$ is the annihilation operator for the cavity $i$, $b_j$ is the annihilation operator of the mechanical mode $j$, and
\begin{equation}
H_d=\sum_{i=1}^3\sum_{j=1}^2\mathcal{E}_{ij}(a_ie^{i(\omega_{d,ij}t+\phi_{ij})}+a_i^{\dagger}e^{-i(\omega_{d,ij}t+\phi_{ij})}),
\end{equation}
describes the Hamiltonian of the pumps with amplitude $\mathcal{E}_{ij}=\mathcal{E}_{ij}^*$, frequency $\omega_{d,ij}$, and phase $\phi_{ij}$.   

We can linearize Hamiltonian (\ref{Ham1}) by expanding the cavity modes around their steady-state field amplitudes, $a_i\rightarrow a_i-\sum_{j=1}^2\alpha_{ij}e^{-i\omega_{d,ij}t}$, where  $|\alpha_{ij}|^2=4|\mathcal{E}_{ij}e^{-i\phi_{ij}}|^2/(\kappa_i^2+4\Delta_{ij}^2)$ is the mean number of photons inside the cavity $i$ induced by the microwave pump due to driving mechanical mode $j$, the $\kappa_i=\kappa_{\mathrm{int},i}+\kappa_{\mathrm{ext},i}$ is the total damping rate of the cavity while $\kappa_{\mathrm{int},i}$ and $\kappa_{\mathrm{ext},i}$ show the intrinsic and extrinsic decay rate of the microwave cavities, respectively. Here,  $\Delta_{ij}=\omega_i-\omega_{d,ij}$ is the detuning of the drive tone with respect to cavity $i$. In the rotating frame with respect to $\sum_{i=1}^{3}\omega_ia_i^{\dagger}a_i+\sum_{i=1}^2(\omega_{m,i}+\delta_{0,i})b_i^{\dagger}b_i$,  the linearized Hamiltonian becomes
\begin{equation}
H=-\sum_{i=1}^2\delta_{0,i}b_i^{\dagger}b_i+\sum_{i=1}^3\Big \{ \Big(\sum_{j=1}^2\big[\alpha_{ij}e^{i\Delta_{ij}t}a_i^{\dagger}+\alpha_{ij}^*e^{-i\Delta_{ij}t}a_i\big]\Big)\Big(\sum_{j=1}^2g_{0,ij}\big[b_je^{-i(\omega_{m,j}+\delta_{0,j})t}+b_j^{\dagger}e^{i(\omega_{m,j}+\delta_{0,j})t}\big]\Big)\Big \}.
\end{equation}
By setting the effective cavity detunings so that $\Delta_{11}=\Delta_{21}=\Delta_{31}=\omega_{m,1}+\delta_{0,1}$ and $\Delta_{12}=\Delta_{22}=\Delta_{32}=\omega_{m,2}+\delta_{0,2}$ and neglecting the terms rotating at $\pm2\omega_{m,1(2)}$ and $\omega_{m,1}+\omega_{m,2}$, the above Hamiltonian reduces to
\begin{equation}\label{Hamfinal}
H=-\sum_{i=1}^2\delta_{0,i}b_i^{\dagger}b_i+\sum_{i=1}^3\sum_{j=1}^2 \Big(G_{ij}a_i^{\dagger}b_j+G_{ij}^*a_ib_j^{\dagger}\Big)+H_{\mathrm{off}}.
\end{equation}
where $G_{ij}=g_{0,ij}|\alpha_{ij}|e^{-i\phi_{ij}}$ is the effective coupling rate between the mechanical mode $j$ and cavity $i$ and $H_{\mathrm{off}}$ describes off-resonant/time dependent interaction between mechanical modes and the cavity fields, and it is given by
\begin{equation}\label{Hoff}
H_{\mathrm{off}}= \sum_{i=1}^3\sum_{j=1}^2\Big[F_{ij}a_i^{\dagger}b_je^{(-1)^{j-1}i\delta \omega_m t}+H.c.\Big]
\end{equation}
where $\delta \omega_m=\omega_{m,2}-\omega_{m,1}+\delta_{0,2}-\delta_{0,1}$ and we define following off-resonant optomechanical coupling parameters
\begin{eqnarray}
F_{11}&=&g_{0,11}|\alpha_{12}|e^{-i\phi_{12}},\,\,F_{12}=g_{0,12}|\alpha_{11}|e^{-i\phi_{11}},\nonumber\\
F_{21}&=&g_{0,21}|\alpha_{22}|e^{-i\phi_{22}},\,\,F_{22}=g_{0,22}|\alpha_{21}|e^{-i\phi_{21}},\\
F_{31}&=&g_{0,31}|\alpha_{32}|e^{-i\phi_{32}},\,\,F_{32}=g_{0,32}|\alpha_{31}|e^{-i\phi_{31}}.\nonumber
\end{eqnarray}

The off-resonant Hamiltonian~(\ref{Hoff}) has an essential role in the nonreciprocity aspect of our device, therefore, it is important to discuss the physical roots of such off-resonant couplings \cite{Buchmann2015,Noguchi2016}. Inspection of Hamiltonians~(\ref{Hamfinal}) and (\ref{Hoff}) reveals that each drive tone generates two different  types of interactions: Resonant coupling in which the drive tone couples a single mechanical mode to a single cavity mode, described by the time-independent part of the Hamiltonian~(\ref{Hamfinal}). Each drive tone also generates an interaction which couples the other mechanical
mode to the cavity off-resonantly. The Hamiltonians~(\ref{Hoff}) explain this off-resonant coupling between cavity fields and mechanical modes. As we will see, these off-resonant couplings alter the mechanical damping rate, which changes the isolation bandwidth and also cools the mechanical modes. In addition, the coupling also introduces mechanical frequency shifts and introduces an effective detuning for the drive tones. Note that, within the rotating wave approximation (RWA) the non-resonant/time-dependent components of the effective linearized interactions can be neglected in the weak coupling regime and when the cavity decay rates $\kappa_i$ are much smaller than the two mechanical frequencies $\omega_{m,i}$  and their difference
\begin{equation}
|F_{ij}|, \kappa_i \ll \omega_{m,j}, |\omega_{m,2}-\omega_{m,1}|.
\end{equation}

Finally, we note that for the isolator case we deal with two cavities coupled two mechanical modes, which mathematically is equivalent to set $G_{31}=G_{32}=F_{31}=F_{32}=0$ in our general model. In this special case, the Hamiltonian~(\ref{Hamfinal}) reduces to the Hamiltonian~(1) presented in the paper
\begin{equation}
H=-\sum_{i=1}^2\delta_{0,i}b_i^{\dagger}b_i+\sum_{i,j=1}^2 \Big(G_{ij}a_i^{\dagger}b_j+G_{ij}^*a_ib_j^{\dagger}\Big)+H_{\mathrm{off}}.
\end{equation}
with 
\begin{equation}\label{Hoff2}
H_{\mathrm{off}}= \sum_{i,j=1}^2\Big[F_{ij}a_i^{\dagger}b_je^{(-1)^{j-1}i\delta \omega_m t}+H.c.\Big].
\end{equation}

%%%%%%%%%%%%%%%%%%%%%

\subsection{Equations of motion and effective model}\label{equationofmotion}

The full quantum treatment of the system can be given in terms of the quantum Langevin equations where we add to the Heisenberg equations the quantum noise acting on the mechanical resonators $b_{\mathrm{in},i}$ with damping rates $\gamma_i$ as well as the cavities’ input fluctuations $a_{\mathrm{in},i}$ with damping rates $\kappa_{\mathrm{ext},i}$. The resulting Langevin equations, including the off-resonate terms, for the cavity modes and mechanical resonators are
\begin{eqnarray}\label{equationmotion1}
\dot a _i&=&-\frac{\kappa_i}{2}a_i-i\sum_{j=1}^2G_{ij}b_j-i\sum_{j=1}^2F_{ij}b_je^{(-1)^{j-1}i\delta \omega_m t}+\sqrt{\kappa_{\mathrm{ext},i}}a_{\mathrm{in},i},\\
\dot b _j&=&\Big(i\delta_{0,j}-\frac{\gamma_{m,j}}{2}\Big)b_j-i\sum_{i=1}^3G_{ij}^*a_i-i\sum_{i=1}^3 F_{ij}^*a_ie^{(-1)^{j}i\delta \omega_m t}+\sqrt{\gamma_{m,j}}b_{\mathrm{in},j},\nonumber
\end{eqnarray}
where $i=1,2,3$ and $j=1,2$. 

In order to study the dynamics of the system we solve the time-dependent quantum Langevin equations~(\ref{equationmotion1}). We use an iterative method to solve these equations by defining a new set of auxiliary operators~(toy modes) and cutting the iteration sequence at higher order dependence to $O(n\,\delta \omega_m;\delta \omega_m^n)$ with $n\geq 2$, which yields
\begin{eqnarray}\label{equationmotion2}
\dot a _i&=&-\frac{\kappa_i}{2}a_i-i\sum_{j=1}^2G_{ij}b_j-i\sum_{j=1}^2F_{ij}B_j+\sqrt{\kappa_{\mathrm{ext},i}}a_{\mathrm{in},i},\nonumber\\
\dot b _1&=&\Big(i\delta_{0,1}-\frac{\gamma_{m,1}}{2}\Big)b_1-i\sum_{i=1}^3G_{i1}^*a_i-i\sum_{i=1}^3 F_{i1}^*A_i^{-}+\sqrt{\gamma_{m,1}}b_{\mathrm{in},1},\nonumber\\
\dot b _2&=&\Big(i\delta_{0,2}-\frac{\gamma_{m,2}}{2}\Big)b_2-i\sum_{i=1}^3G_{i2}^*a_i-i\sum_{i=1}^3 F_{i2}^*A_i^{+}+\sqrt{\gamma_{m,2}}b_{\mathrm{in},2},\\
\dot A_i^{+}&=&(i\delta \omega_m-\frac{\kappa_i}{2})A_i^{+}-i\Big(F_{i2}b_2+G_{i1}B_1\Big),\nonumber\\
\dot A_i^{-}&=&-(i\delta \omega_m+\frac{\kappa_i}{2})A_i^{-}-i\Big(F_{i1}b_1+G_{i2}B_2\Big),\nonumber\\
\dot B_1&=&\Big(i[\delta \omega_m+\delta_{0,1}]-\frac{\gamma_{m,1}}{2}\Big)B_1-i\sum_{i=1}^3\Big(F_{i1}^*a_i+G_{i1}^*A_i^{+}\Big),\nonumber\\
\dot B_2&=&-\Big(i[\delta \omega_m-\delta_{0,2}]+\frac{\gamma_{m,2}}{2}\Big)B_2-i\sum_{i=1}^3\Big(F_{i2}^*a_i+G_{i2}^*A_i^{-}\Big),\nonumber
\end{eqnarray}
where $i=1,2,3$. The auxiliary modes~ $A_{i}^{\pm}=a_ie^{\pm i\delta \omega_mt},\, B_1=b_1e^{ i\delta \omega_mt}$ and $B_2=b_2e^{-i\delta \omega_mt}$ describe the off-resonant components of the equations of motion. Here, we take $\delta \omega_m$ to be much larger than the relevant system frequencies, i.e. $ \delta \omega_m \gg \gamma_{m,i},\delta_{0,i},\omega $,  and can thus adiabatically
eliminate the auxiliary modes by taking~$\dot B_j=\dot A_i^{\pm}=0$ in Eqs.~(\ref{equationmotion2}), which results in the following equations for the auxiliary modes
\begin{eqnarray}\label{equationmotion4}
A_i^{+}&=&\frac{i\Big(F_{i2}b_2+G_{i1}B_1\Big)}{(i\delta \omega_m-\frac{\kappa_i}{2})},\nonumber\\
A_i^{-}&=-&\frac{i\Big(F_{i1}b_1+G_{i2}B_2\Big)}{(i\delta \omega_m+\frac{\kappa_i}{2})},\\
B_1&=&\frac{i\sum_{i=1}^3\Big(F_{i1}^*a_i+G_{i1}^*A_i^{+}\Big)}{\Big(i[\delta \omega_m+\delta_{0,1}]-\frac{\gamma_{m,1}}{2}\Big)},\nonumber\\
B_2&=&-\frac{i\sum_{i=1}^3\Big(F_{i2}^*a_i+G_{i2}^*A_i^{-}\Big)}{\Big(i[\delta \omega_m-\delta_{0,2}]+\frac{\gamma_{m,2}}{2}\Big)},\nonumber
\end{eqnarray}
In the limit of $\delta \omega_m \rightarrow \infty$, the contribution of all auxiliary modes can be totally neglected in the dynamics of the system, i.e. $ \{B_j,A_i^{\pm}\}\rightarrow 0$. In this case the off-resonant interactions between the mechanical modes and cavities are negligible and we can safely ignore the time-dependent components of the Hamiltonian (i.e. $H_{\mathrm{off}}=0$). However, in our system due to finite value of $\delta \omega_m \approx \kappa_i/2$, we cannot ignore these off-resonant interactions.

We can simply further the equations of motion for the main modes by substituting Eqs.~(\ref{equationmotion4}) into the equations of motion for $a_i$ and $b_j$ in Eqs.~(\ref{equationmotion2}) and assuming $\delta \omega_m,\kappa_i \gg \Big\{ |\delta_{0,j}|,\gamma_{m,j},|G_{ij}|,|F_{ij}| \Big\}$,  
\begin{eqnarray}\label{Mainequation}
\dot a _i & \approx & -\frac{\kappa_i}{2}a_i-i\sum_{j=1}^2G_{ij}b_j+\sqrt{\kappa_{\mathrm{ext},i}}a_{\mathrm{in},i},\nonumber\\
\dot b _1 & \approx & \Big(i\delta_{1}-\frac{\Gamma_{m,1}}{2}\Big)b_1-i\sum_{i=1}^3G_{i1}^*a_i+\sqrt{\gamma_{m,1}}b_{\mathrm{in},1},\\
\dot b _2 & \approx & \Big(i\delta_{2}-\frac{\Gamma_{m,2}}{2}\Big)b_2-i\sum_{i=1}^3G_{i2}^*a_i+\sqrt{\gamma_{m,2}}b_{\mathrm{in},2},\nonumber
\end{eqnarray}
where $\delta_{j}$ and $\Gamma_{m,j}$ are the effective detuning and damping rates of the mechanical modes, respectively, and they are given by
\begin{eqnarray}
\delta_{1}&=& \delta_{0,1}+\delta \omega_m \sum_{i=1}^3 \frac{4|F_{i1}|^2}{4\delta \omega_m^2+\kappa_i^2},\nonumber\\
\delta_{2}&=& \delta_{0,2}-\delta \omega_m \sum_{i=1}^3 \frac{4|F_{i2}|^2}{4\delta \omega_m^2+\kappa_i^2},\\
\Gamma_{m,1}&=& \gamma_{m,1}+\sum_{i=1}^3 \frac{4\kappa_i|F_{i1}|^2}{4\delta \omega_m^2+\kappa_i^2},\nonumber\\
\Gamma_{m,2}&=& \gamma_{m,2}+\sum_{i=1}^3 \frac{4\kappa_i|F_{i2}|^2}{4\delta \omega_m^2+\kappa_i^2}.\nonumber
\end{eqnarray}
Note that in the derivation of Eqs.~(\ref{Mainequation}) we assume that the off-resonant interaction does not considerably modify the self-interaction and damping rate of the cavity modes.  
Inspection of Eqs.~(\ref{Mainequation}) reveals that the off-resonant coupling between mechanical modes and cavities shifts the resonance frequency and damps/cools the mechanical modes by introducing a cross-damping between them. The strength of the frequency shift and the cross-damping is given by the off-resonant optomechanical coupling parameters $F_{ij}$, which indicates that the drive tones creates an effective coupling between the two mechanical modes. In the weak coupling regime and for very large $\delta \omega_m$ this cross-coupling is negligible, thus $\delta_{j}\approx \delta_{0,j}$ and $\Gamma_{m,j}\approx \gamma_{m,j}$. 

We can solve the  Eqs.~(\ref{Mainequation}) in the Fourier domain to obtain the microwave cavities' variables. Eliminating the mechanical degrees of freedom from the equations of motion~(\ref{Mainequation}) and writing the remaining equations in the matrix form, we obtain
\begin{eqnarray}\label{equationmatrix}
\Big(\textbf{M}-i\omega \textbf{I}\Big)\left(
  \begin{array}{cc}
    a_1 \\
      a_2\\
      a_3\\
  \end{array}
\right)=\left(
  \begin{array}{cc}
    \sqrt{\kappa_{\mathrm{ext},1}}a_{\mathrm{in},1}-iG_{11}\chi_{m,1}(\omega)\sqrt{\gamma_{m,1}}b_{\mathrm{in},1}-iG_{12}\chi_{m,2}(\omega)\sqrt{\gamma_{m,2}}b_{\mathrm{in},2}\\
          \sqrt{\kappa_{\mathrm{ext},2}}a_{\mathrm{in},2}-iG_{21}\chi_{m,1}(\omega)\sqrt{\gamma_{m,1}}b_{\mathrm{in},1}-iG_{22}\chi_{m,2}(\omega)\sqrt{\gamma_{m,2}}b_{\mathrm{in},2}\\
            \sqrt{\kappa_{\mathrm{ext},3}}a_{\mathrm{in},3}-iG_{31}\chi_{m,1}(\omega)\sqrt{\gamma_{m,1}}b_{\mathrm{in},1}-iG_{32}\chi_{m,2}(\omega)\sqrt{\gamma_{m,2}}b_{\mathrm{in},2}\\
  \end{array}
\right),\nonumber\\
\end{eqnarray}
where $\chi_j^{-1}(\omega)=\Gamma_{m,j}/2-i(\omega+\delta_{j})$ is the mechanical susceptibility for mode $j$ and we introduced the drift matrix 
\begin{eqnarray}\label{denm}
\textbf{M}=\left(
  \begin{array}{ccc}
    \frac{\kappa_1}{2}+\chi_{m,1}(\omega)|G_{11}|^2+\chi_{m,2}(\omega)|G_{12}|^2 & \chi_{m,1}(\omega)G_{11}G_{21}^*+\chi_{m,2}(\omega)G_{12}G_{22}^*& \chi_{m,1}(\omega)G_{11}G_{31}^*+\chi_{m,2}(\omega)G_{12}G_{32}^*\\
   \chi_{m,1}(\omega)G_{11}^*G_{21}+\chi_{m,2}(\omega)G_{12}^*G_{22}& \frac{\kappa_2}{2}+\chi_{m,1}(\omega)|G_{21}|^2+\chi_{m,2}(\omega)|G_{22}|^2&  \chi_{m,1}(\omega)G_{31}^*G_{21}+\chi_{m,2}(\omega)G_{32}^*G_{22}\\
             \chi_{m,1}(\omega)G_{11}^*G_{31}+\chi_{m,2}(\omega)G_{12}^*G_{32}&  \chi_{m,1}(\omega)G_{21}^*G_{31}+\chi_{m,2}(\omega)G_{22}^*G_{32}&\frac{\kappa_3}{2}+\chi_{m,1}(\omega)|G_{31}|^2+\chi_{m,2}(\omega)|G_{32}|^2 \\
  \end{array}
\right).\nonumber
\end{eqnarray}

By substituting the solutions of Eq.~(\ref{equationmatrix}) into the corresponding input-output formula for the cavities’ variables, i.e. $a_{\mathrm{out},j}=\sqrt{\kappa_{\mathrm{ext},j}}a_j-a_{\mathrm{in},j}$, we obtain
\begin{eqnarray}\label{equationmatrix2}
\left(
  \begin{array}{cc}
    a_{\mathrm{out},1} \\
       a_{\mathrm{out},2} \\
    a_{\mathrm{out},3} \\
  \end{array}
\right)=\textbf{T}.\Big(\textbf{M}-i\omega \textbf{I}\Big)^{-1}.\left(
  \begin{array}{cc}
    \sqrt{\kappa_{\mathrm{ext},1}}a_{\mathrm{in},1}-iG_{11}\chi_{m,1}(\omega)\sqrt{\gamma_{m,1}}b_{\mathrm{in},1}-iG_{12}\chi_{m,2}(\omega)\sqrt{\gamma_{m,2}}b_{\mathrm{in},2}\\
          \sqrt{\kappa_{\mathrm{ext},2}}a_{\mathrm{in},2}-iG_{21}\chi_{m,1}(\omega)\sqrt{\gamma_{m,1}}b_{\mathrm{in},1}-iG_{22}\chi_{m,2}(\omega)\sqrt{\gamma_{m,2}}b_{\mathrm{in},2}\\
            \sqrt{\kappa_{\mathrm{ext},3}}a_{in,3}-iG_{31}\chi_{m,1}(\omega)\sqrt{\gamma_{m,1}}b_{\mathrm{in},1}-iG_{32}\chi_{m,2}(\omega)\sqrt{\gamma_{m,2}}b_{\mathrm{in},2}\\
  \end{array}
\right)-\left(
  \begin{array}{cc}
    a_{\mathrm{in},1} \\
       a_{\mathrm{in},2} \\
    a_{\mathrm{in},3} \\
  \end{array}
\right),\nonumber\\
\end{eqnarray}
where we defined $\textbf{T}=\mathrm{Diag}\big[\sqrt{\kappa_{\mathrm{ext},1}},\sqrt{\kappa_{\mathrm{ext},2}},\sqrt{\kappa_{\mathrm{ext},3}}\big]$.

\subsection{Scattering matrix and nonreciprocity for a two-port device}\label{ISOsection}
In this section, we verify the details of our analysis in the isolator section of the main paper and we examine our model to see how the nonreciprocity arises in a two-port electromechanical system. Here, we are only interested in the response an electromechanical system comprised of two microwave cavities and two mechanical modes. Therefore, by setting $G_{3j}\rightarrow 0$ and $\delta_{1}=-\delta_{2}=\delta$ in Eq.~(\ref{equationmatrix2}) and assuming $\phi_{22}=\phi,\,\phi_{11}=\phi_{12}=\phi_{21}=0$, we can find the ratio of backward to forward transmission
\begin{eqnarray}\label{nonreciprocitySI}
\lambda:=\frac{S_{12}(\omega)}{S_{21}(\omega)}=\frac{\sqrt{C_{11}C_{21}}\Sigma_{m,2}(\omega)+\sqrt{C_{12}C_{22}}\Sigma_{m,1}(\omega)e^{i\phi}}{\sqrt{C_{11}C_{21}}\Sigma_{m,2}(\omega)+\sqrt{C_{12}C_{22}}\Sigma_{m,1}(\omega)e^{-i\phi}},
\end{eqnarray}
as specified in Eq.~(2) of the paper. Here, $\Sigma_{m,j}=1+2i\big[(-1)^j\delta-\omega\big]/\Gamma_{m,j}$ is the inverse of the mechanical susceptibility divided by the effective mechanical linewidth $\Gamma_{m,j}$. Examination of Eq.~(\ref{nonreciprocitySI}) shows that the nominator and denominator of this equation are not equal and they possess different relative phase. This asymmetry is the main source of the nonreciprocity and appearance of isolation in the system. In particular, at a phase
\begin{equation}\label{phaseeq}
e^{i\phi}=-\sqrt{\frac{C_{11}C_{21}}{C_{12}C_{22}}}\frac{\Sigma_{m,2}(\omega)}{\Sigma_{m,1}(\omega)},
\end{equation}
the nominator of the Eq.~(\ref{nonreciprocitySI}) will be zero, therefore,  backward transmission $S_{12}$ is canceled while forward transmission $S_{21}$ is non-zero. Rewriting Eq.~(\ref{phaseeq}) gives
\begin{equation}\label{tanphase}
\mathrm{tan}[\phi(\omega)]=\frac{\delta(\Gamma_{m,1}+\Gamma_{m,2})+\omega(\Gamma_{m,2}-\Gamma_{m,1})}{\Gamma_{m,1}\Gamma_{m,2}/2-2(\delta^2-\omega^2)}. 
\end{equation}
By neglecting the contribution of the off-resonant term in the response of the system, i.e. $\Gamma_{m,j}\rightarrow \gamma_{m,j}$  the Eq.~(\ref{tanphase}) reduces to Eq.~(3) of the paper. At the optimum phase (\ref{phaseeq}) and at cavity resonance, the transmission in the forward direction is given by
\begin{equation}\label{trans11}
S_{21}=-
\frac{2\sqrt{\eta_1\eta_2}\big[\Sigma_{m,1}(0)\Sigma_{m,2}(0)\big]\big(\sqrt{C_{11}C_{21}}\Sigma_{m,2}(0)+\sqrt{C_{12}C_{22}}\Sigma_{m,1}(0)e^{-i\phi}\big)}{\big[C_{11}\Sigma_{m,2}(0)+C_{12}\Sigma_{m,1}(0)+\Sigma_{m,1}(0)\Sigma_{m,2}(0)\big]\big[C_{21}\Sigma_{m,2}(0)+C_{22}\Sigma_{m,1}(0)+\Sigma_{m,1}(0)\Sigma_{m,2}(0)\big]}.\nonumber
\end{equation}
For equal mechanical damping $\Gamma_{m,1}=\Gamma_{m,2}=\Gamma$ (equivalent to $\gamma_{m,1}=\gamma_{m,2}=\gamma$ of the main text) and at equal cooperativities for all four optomechanical couplings~($C_{ij}=\mathcal{C}$) the above equation reduces to

\begin{eqnarray}\label{trans11}
S_{21}=-\sqrt{\eta_1\eta_2}\Big[
\frac{4i\,\delta(1-2i\delta/\Gamma)}{\mathcal{C}\Gamma(1+\frac{1+4\delta^2/\Gamma^2}{2\mathcal{C}})^2}\Big]
\end{eqnarray}
as specified in Eq.~(4) of the paper. For the particular cooperativity $ 2\mathcal{C}=1+4\delta^2/\Gamma^2$, the power transmission in forward direction is given by
\begin{equation}
|S_{21}|^2=\eta_1 \eta_2\Big(1-\frac{1}{2\mathcal{C}}\Big).
\end{equation}
By neglecting the off-resonant interaction all damping rates reduce to $\Gamma_{m,j}\approx \gamma_{m,j}$ which is consistent with our notation in the main text.   We also note that the frequency shifts due to off-resonant interaction for the isolator system discussed in the main text are given by $(\delta_1,\delta_2)/2\pi=(-84,233)$~Hz while the cross-damping rates are $(\Gamma_{m,1},\Gamma_{m,2})/2\pi=(190,407)$~Hz.

\subsection{Theoretical model for the circulator}
The theoretical model, we presented in Eqs.~(\ref{equationmotion2}), or equivalently Eq.~(\ref{equationmatrix2}), fully describes the nonreciprocal behavior of the system for the case of the circulator. In order to check this, in Fig.~\ref{circulatorSI} we show both measured experimental data and the theoretical prediction. The theoretical model is in excellent agreement with the experiment and can perfectly describe the nonreciprocity of photon transmission for both forward and backward circulation. 
\begin{figure}[t]
\centering
\includegraphics[width=0.6\columnwidth]{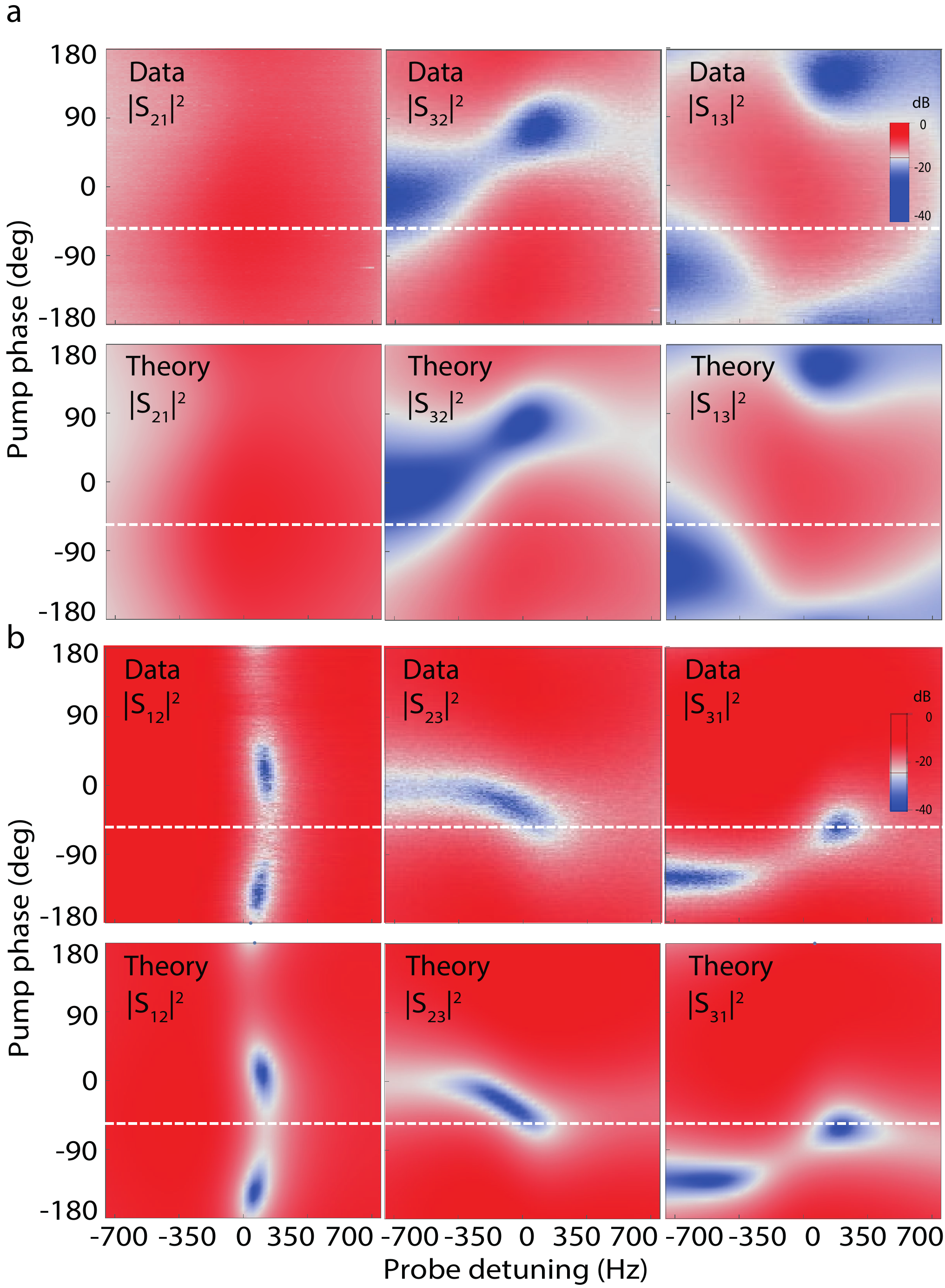}
\caption{\textbf{Full scattering parameters of the circulator.} \textbf{a}, Measured power transmission and theoretical model in forward direction ($|S_{21}|^2$, $|S_{32}|^2$, and $|S_{13}|^2$) as a function of detuning and pump phase. \textbf{b}, Measured power transmission and theoretical model in backward direction ($|S_{12}|^2$, $|S_{23}|^2$, and $|S_{31}|^2$) as a function of detuning and pump phase.} \label{circulatorSI}
\end{figure}

\section{Added noise}\label{addnoise}
In this section, we discuss the noise properties of the system and present data for the added noise during the frequency conversion when operated as a circulator. 

Equation~(\ref{equationmatrix2}) explains that due to the linear nature of the input-output theorem and in the absence of the input coherent signal, the output of each cavity is a linear combination of the electromagnetic input noise $a_{\text{in},i}$ and mechanical noise $b_{\text{in},j}$. Therefore, Eq.~(\ref{equationmatrix2}) can be rewritten in the following general form
\begin{equation}\label{outputop}
a_{\mathrm{out},i}=\sum_{j=1}^{3}S_{i,j}a_{\mathrm{in,j}}+\sum_{j=1}^{2}T_{i,j}b_{\mathrm{in,j}},
\end{equation}
where $S_{i,j}$ and $T_{i,j}$ are the scattering matrices. 
Operating under the white noise assumption, the zero-mean quantum fluctuations $a_{\text{in},i} $ and $b_{\text{in},j}$ satisfy the correlations $\langle O_{\text{in},i/j}(t)O_{\text{in},i/j}^\dagger(t^\prime)\rangle=(\bar{N}_{i/j}+1)\delta(t-t^\prime)$, $\langle O_{\text{in},i/j}^\dagger(t) O_{\text{in},i/j}(t^\prime)\rangle=\bar{N}_{i/j}\delta(t-t^\prime)$, and $\langle O_{\text{in},i/j}(t)O_{\text{in},i/j}(t^\prime)\rangle=0$ where $i=1,2,3$ for $O=a$, and $j=1,2$ for $O=b$) and $\bar{N}_{i/j}=1\big/\bigl\{\exp\bigl[\hbar\omega_i/(k_\text{B}T_i)\bigr]-1\bigr\}$ ($\bar{N}_{m,j}=1\big/\bigl\{\exp\bigl[\hbar\omega_{m,j}/(k_\text{B}T_j)\bigr]-1\bigr\}$) are the thermal photon (phonon) occupancies of the cavities (mechanical resonator) for $i=1,2,3$ ($j=1,2$) at temperature $T_i$. The output of the cavities are then sent through a chain of amplifiers. The electromagnetic modes at the output of the amplifiers are given by
\begin{equation}\label{outputop1}
A_{\mathrm{out},i}=\big(\sqrt{G_i}a_{\mathrm{out},i}+\sqrt{G_i-1}c^{\dagger}_{\mathrm{amp},i}\big),
\end{equation}
where $G_i$ is the effective gain of the amplifier chain at port $i$ and $c_{\mathrm{amp},i}$ is the added noise operator of the amplifiers. We can now
write the expression for the single sided power spectral density as measured by a spectrum analyzer, in the
presence of all relevant noise sources
\begin{equation}\label{noise4}
S_{\mathrm{noise},i}(\omega)=\hbar \omega \int_{-\infty}^{\infty}d \omega' \langle A_{\mathrm{out},i}^{\dagger}(\omega')A_{\mathrm{out},i}(\omega')\rangle. 
\end{equation}

Substituting Eqs.~(\ref{outputop}) and (\ref{outputop1}) into Eq.~(\ref{noise4}), assuming $G_i\approx G_i-1=10^{\mathcal{G}_i/10}$ where $\mathcal{G}_i$ is the gain in dB, and using the white correlation functions for the noise operators, we find
\begin{equation}\label{noise}
S_{\mathrm{noise},i}(\omega)=\hbar \omega 10^{\mathcal{G}_i/10} (1+n_{\mathrm{amp},i}+n_{\mathrm{add},ij}),
\end{equation}
where $n_{\mathrm{amp},ij}$ is the total noise added by the amplifier chains and $n_{\mathrm{add},i}$ is the total noise added by the cavities and mechanical resonators associated with the photon conversion from cavity $j$ to cavity $i$. 

Measuring the output noise spectrum and having calibrated the gain of the amplifiers at each port~$(\mathcal{G}_1,\mathcal{G}_2,\mathcal{G}_3)=(67.5,64,60.5)$~dB, we can accurately infer the amplifiers added noise quanta at each port $(n_{\mathrm{amp},1},n_{\mathrm{amp},2},n_{\mathrm{amp},3})=(23,23,33)\pm2$. The only remaining unknown parameter in Eq.~(\ref{noise}) is $n_{\mathrm{add},ij}$ which can be found by measuring the noise properties of the three cavities when all six pumps are on and compare them to the case when the pumps are off. In the Fig.~\ref{noisefig} we show the measured added noise photons for all six transmission parameters of the circulator. On resonance where the directionality is maximized we find $(n_{\mathrm{add},21},n_{\mathrm{add},32},n_{\mathrm{add},13})=(4,6.5,3.6)$ in the forward direction and $(n_{\mathrm{add},12},n_{\mathrm{add},23},n_{\mathrm{add},31})=(4,4,5.5)$ in the backward direction. 

\begin{figure}[t]
\centering
\includegraphics[width=0.7\columnwidth]{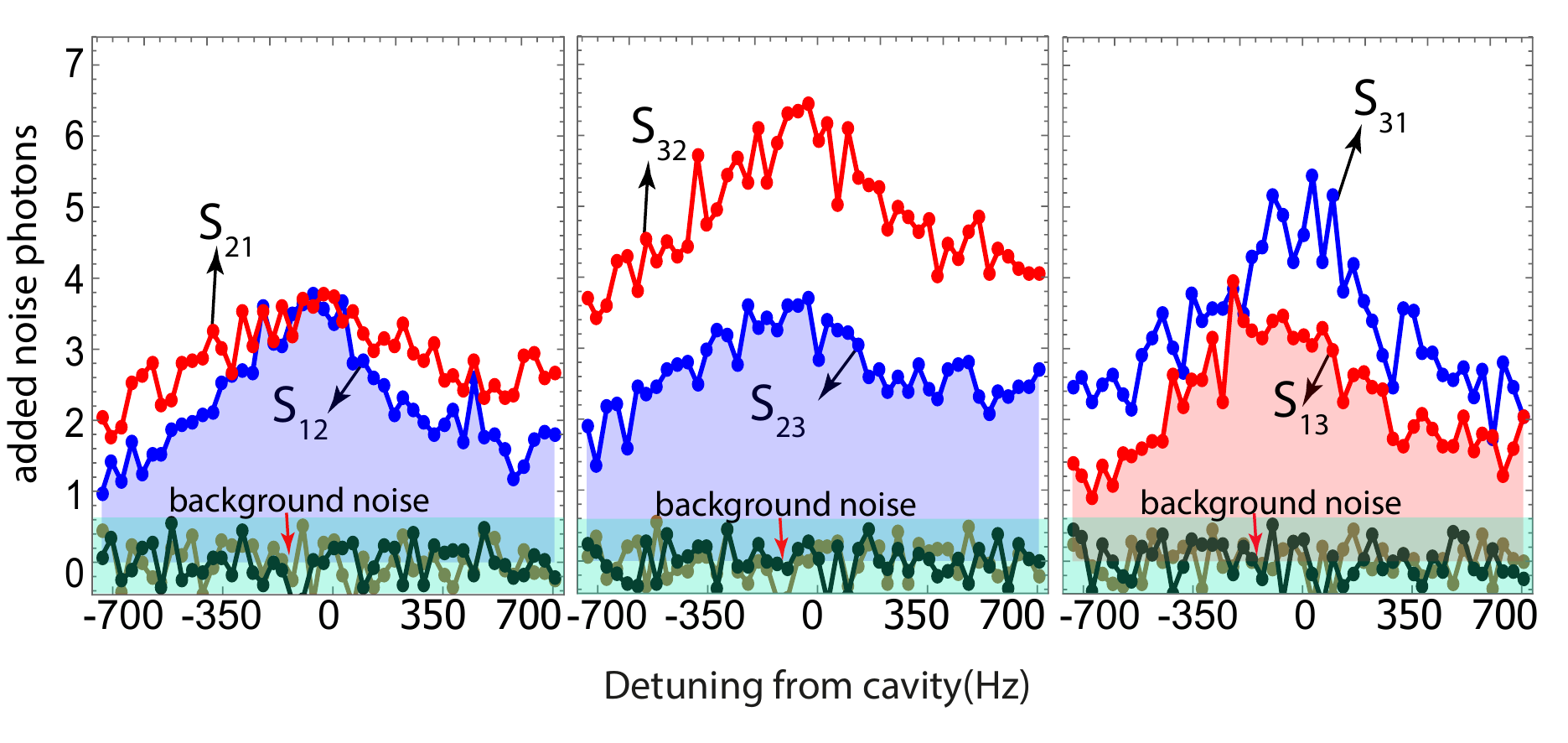}
\caption{\textbf{Added noise photons of the circulator.} Measured circulator noise properties in forward direction ($|S_{21}|^2$, $|S_{32}|^2$, and $|S_{13}|^2$ in red) and backward direction ($|S_{12}|^2$, $|S_{23}|^2$, and $|S_{31}|^2$ in blue) as well as the measured background noise as a function of detuning.} \label{noisefig}
\end{figure}

\end{widetext}
\newpage

\end{document}